\documentclass[12pt]{article}
\usepackage{epsfig}
\usepackage{graphicx}

\def\beq{\begin{equation}}
\def\eeq#1{\label{#1}\end{equation}}
\def\eeqn{\end{equation}}

\def\beqa{\begin{eqnarray}}
\def\eeqa#1{\label{#1}\end{eqnarray}}
\def\eeqan{\end{eqnarray}}

\let\bar=\overbar

\def\Dslash{\not{\hbox{\kern-4pt $D$}}}
\def\dslash{\not{\hbox{\kern-2pt $\del$}}}

\def\msb{{\bar{\ssstyle M \kern -1pt S}}}

\usepackage{fancyhdr,graphicx}
\def\Title#1{\begin{center} {\Large {\bf #1} } \end{center}}

\begin{document}

\Title{Equation of State in a Generalized Relativistic Density Functional Approach}

\bigskip\bigskip


\begin{raggedright}

{\it 
Stefan Typel\\
\bigskip
GSI Helmholtzzentrum f\"{u}r Schwerionenforschung GmbH, 
Theorie,
Planckstra\ss{}e 1,
64291 Darmstadt, Germany
}

\end{raggedright}

\section{Introduction}

In many simulations of astrophysical objects and phenomena, the equation of state (EoS)
of dense matter is an essential ingredient. It determines, e.g., the dynamical evolution 
of core-collapse supernovae \cite{Janka:2006fh,Janka:2012wk} 
and neutron star mergers \cite{Bauswein:2014qla}, and
the structure of compact stars \cite{Lattimer:2006xb}. 
The application of an EoS is reasonable if
the timescales of reactions are much smaller than those of the system evolution
and thermodynamic equilibrium can be assumed to hold. In general, a global EoS is required
that covers a wide range in temperature, density and isospin asymmetry.
These conditions affect the chemical composition of matter 
and the nucleosynthesis. 

A critical examination of existing global EoS models 
\cite{compose} 
suggests that the development of an
improved EoS is worthwhile. The set of constituent particles should be enlarged 
considerably including not only nucleons, charged leptons and photons but
also a ``complete'' table of nuclei, mesons, hyperons or even quarks as degrees of
freedom at high densities and temperatures. The model parameters have to be contrained
better taking, e.g., properties of nuclei, results of heavy-ion collisions or
compact star observations into account. Correlations should be considered more seriously, e.g.,
at low-densities where the virial equation of state (VEoS), which is determined by nucleon-nucleon
correlations, is a model-independent benchmark
\cite{Horowitz:2005nd,O'Connor:2007eb}.
For composite particles such as nuclei
the dissolution in the medium (Mott effect) has to be described properly
\cite{RoMuSchu1982,RoSchmMuSchu1983}. 
Electromagnetic correlations are essential in order to model the solidification/melting at low
temperatures. Phase transitions and the appearance of 'non-congruent' features
have to be treated correctly \cite{Hempel:2013tfa} with non-negligible differences
between nuclear matter and stellar matter. Obviously, it is a tremendous challenge
to cover the full range of thermodynamic variables in a single unified model.

Information on correlations are encoded in spectral functions, which have a complicated
structure in general. Often, a quasiparticles (QP) approach is employed as an approximation.
The QP properties change inside the medium and the size of residual correlations is reduced.
The QP concept is very successful in nuclear physics, e.g., in phenomenological
mean-field models (Skyrme, Gogny, relativistic) or the treatment of pairing correlations
using a Bogoliubov transformation \cite{Bender:2003jk}.
In the ultimate limit, an exact diagonalisation of the
Hamiltonian of the interacting many-body system leads to a system of independent
QP that can be many-body states. At low densities, clusters appear as new degrees of freedom
as described in the VEoS. In order to consider these features, a 
generalized relativistic density functional (gRDF) was developed. It takes the correct limits
and explicit cluster degrees of freedom into account.

\section{Generalized relativistic density functional}

The gRDF model 
\cite{Typel:2009sy,Voskresenskaya:2012np,Typel:2012esa,Typel:2013zna}
is based on a grand canonical approach. It is an extension of 
a conventional relativistic mean-field model with density dependent couplings
\cite{Typel:1999yq}.
All thermodynamic quantities are derived from a grand canonical potential density
$\omega(T,\{\mu_{i}\})$, which depends on the temperature $T$ and the set of chemical
potentials $\mu_{i}$ of all particles. The present set of particle species comprises
baryons (nucleons and hyperons), nuclei, charged leptons and photons.
Besides light nuclei (${}^{2}$H, ${}^{3}$H, ${}^{3}$He, ${}^{4}$He) a full table
of heavy nuclei ($^{A}Z$ with $A>4$, $N,Z \leq 184$) is included, too. 
Experimental binding energies
are used or, if not available, predictions from the DZ model
\cite{Duflo:1994tr}. 
Internal excitations of heavy nuclei are considered with temperature dependent
degeneracy factors obtained with appropriate level densities.
Effective continuum resonances represent nucleon-nucleon scattering correlations
and ensure the correct low density limit in accordance with the VEoS
\cite{Voskresenskaya:2012np}.

All massive particles are treated as QP with scalar ($S_{i}$) and vector ($V_{i}$) potentials.
The effective interaction is modeled by an exchange of mesons ($\sigma$, $\omega$, $\rho$)
with density-dependent couplings to the nucleons, both free and bound in nuclei, using
the well constrained DD2 parametrization \cite{Typel:2009sy}. 
It gives very reasonable nuclear matter
parameters at a saturation density of $n_{\rm sat}=0.149$~fm$^{-3}$, 
such as a binding energy per nucleon $E/A=16.02$~MeV, 
a compressibility $K=242.7$~MeV, a symmetry energy $J=31.67$~MeV 
and a slope parameter $L=55.04$~MeV.
The neutron matter EoS lies within the error bounds of recent
chiral effective field theoretical calculations
\cite{Tews:2012fj,Kruger:2013kua}.
Both potentials $S_{i}$ and $V_{i}$ receive contributions from the meson fields.
For composite particles, the scalar potential contains an additional
mass shift $\Delta m_{i}$ that depends on all particle densities and temperature.
It mainly takes the blocking of states by the Pauli exclusion principle into account
and serves to describe the dissolution of clusters by reducing the particle binding energy.
This microscopically motivated approach replaces the traditional, purely geometric
concept of the excluded-volume mechanism \cite{Hempel:2011kh}. 
The vector potential $V_{i}$ includes a 
``rearrangment'' contribution due to the density dependence of the meson-nucleon couplings, which
is required for the thermodynamic consistency of the model,
and an electromagnetic correction to account for electron sceening effects in
stellar matter. 



\section{Symmetry energy and neutron skins of nuclei}

The isospin dependence of the effective interaction in the gRDF model determines
the density dependence of the symmetry energy. It is crucial for a proper description
of the structure of neutron stars, see, e.g., the topical issue
on the symmetry energy \cite{Li:2014oda}.
A strong correlation of the neutron skin thickness
$\Delta r_{np}$ of heavy nuclei
with the slope of the neutron matter equation of state 
\cite{Brown:2000pd,TyBr2001}
or the slope parameter $L$
of the symmetry energy is observed when the predictions of a large number 
of mean-field calculations, both relativistic and non-relativistic,
are compared, see, e.g., \cite{Vinas:2013hua}.
In recent years, many attempts were made to determine the symmetry energy at
saturation $J$ and the parameter $L$ from experiments, e.g., by measuring 
the neutron skin thickness 
of ${}^{208}$Pb and using the $\Delta r_{np}$ vs.\ $L$ correlation. 
Since the calculations of neutron skin thicknesses are based on mean-field models, 
the question arises whether few-nucleon correlations can effect the results.

The gRDF approach can be employed to describe the formation of nuclei inside matter
at finite temperatures by using an extended Thomas-Fermi approximation
in spherical Wigner-Seitz cells \cite{Typel:2012esa}. In a calculation with
nucleons and light clusters as degrees of freedom,
it is observed that the probability of finding light clusters 
is enhanced at the surface of the heavy nucleus as compared to the surrounding
low-density gas. The gRDF model can be extended to the description of
heavy nuclei in vacuum at zero temperature to study cluster correlations.
In this case, only the
$\alpha$-particle remains as the relevant light cluster. Its density distribution is
obtained from the $\alpha$-particle ground state wave function that is calculated
self-consistently in the WKB approximation. For the chain of Sn nuclei, a distinct
reduction of the neutron skin thickness is observed when $\alpha$-particle correlations
are considered \cite{Typel:2014tqa}. 
However, the effect vanishes for very neutron-rich nuclei or for
nuclei with roughly the same neutron and proton numbers without a neutron skin.
A variation of the isovector dependent part of the effective interaction allows
to study the $\Delta r_{np}$ vs.\ $L$ correlation, e.g., for a ${}^{208}$Pb nucleus.
A systematic shift is observed that might affect the determination of the slope parameter
$L$ from measurements of the neutron skin thickness, at least as a systematic error.
It is envisaged to investigate experimentally 
the predicted formation of $\alpha$-particles at the surface of Sn nuclei 
in quasi-elastic (p,p$\alpha$) reactions at RCNP, Osaka \cite{Aumann2014}.

\section{Outlook}

The present version of the gRDF model includes only hadronic and leptonic
degrees of freedom where nuclei are described as clusters composed
of nucleons. At high densities or temperatures a phase transition to
quark matter is expected. Hence, quark degrees of freedom should be
incorporated into the approach. On the other hand, at low densities and temperatures,
quarks should be confined in nucleons. In a preliminary extension of the gRDF model
with quarks, a phenomenological description of confinement will be implemented.
The idea is to apply an ``inverse'' excluded-volume approach that permits
the quarks to propagate freely only above a certain (scalar) density of the system.
For this purpose, the classical excluded-volume mechanism is generalized
by allowing more general dependencies of the ``available volume fraction''.
The correct quantum statistics and a relativistic description are considered, too. 
The relevant theoretical formulation to guarantee the thermodynamic consistency 
of the approach has been developed and exploratory calculations have been preformed.

Another extension of the gRDF model concerns the introduction of
more general meson-nucleon couplings in the Lagrangian density. 
In conventional RMF approaches with density-dependent
couplings, the nucleon self-energies only depend on densities.
As known from Dirac-Brueckner calculations of nuclear matter, 
they should also depend on the nucleon momentum or energy. This dependence
can be mapped to modified effective density dependent
meson-nucleon couplings \cite{Hofmann:2000vz}, but the full dependence
should be kept in order to comply with the optical
potential constraint at high nucleon energies. This can be achieved
in a RMF model with density-dependent and non-linear derivative meson-nucleon
couplings of general functional form \cite{Antic:2015tga}. 
Preliminary studies indicate a softening of the EoS at high densities, however,
for a reliable fit of the model parameters, the approach has to be applied
to the description of finite nuclei. Work in this direction is in progress.


\subsection*{Acknowledgement}

The author thanks Sofija Anti\'{c}, David Blaschke, Jaroslava Hrt\'{a}nkov\'{a}, 
Thomas Kl\"{a}hn, Gevorg Poghosyan, Gerd R\"{o}pke, Maria Voskresenskaya and
Hermann Wolter for the collaboration, discussions and encouragement 
during various stages in development of the gRDF model and extensions.
This work was supported by the Helmholtz Association (HGF) through the 
Nuclear Astrophysics Virtual Institute (VH-VI-417). The participation of
the author at the CSQCD IV workshop was made possible by
NewCompStar, COST Action MP1304.


\begin{thebibliography}{99}

\bibitem{Janka:2006fh}
  H.~T.~Janka, K.~Langanke, A.~Marek, G.~Martinez-Pinedo and B.~Mueller,
  Phys.\ Rept.\  {\bf 442} (2007) 38
  [astro-ph/0612072].

\bibitem{Janka:2012wk}
  H.~T.~Janka,
  Ann.\ Rev.\ Nucl.\ Part.\ Sci.\  {\bf 62} (2012) 407
  [arXiv:1206.2503 [astro-ph.SR]].

\bibitem{Bauswein:2014qla}
  A.~Bauswein, N.~Stergioulas and H.-T.~Janka,
  Phys.\ Rev.\ D {\bf 90} (2014) 2,  023002
  [arXiv:1403.5301 [astro-ph.SR]].

\bibitem{Lattimer:2006xb}
  J.~M.~Lattimer and M.~Prakash,
  Phys.\ Rept.\  {\bf 442} (2007) 109
  [astro-ph/0612440].



\bibitem{compose}
  CompStar Online Supernovae Equations of State, {\tt http://compose.obspm.fr/}

\bibitem{Horowitz:2005nd}
  C.~J.~Horowitz and A.~Schwenk,
  Nucl.\ Phys.\ A {\bf 776} (2006) 55
  [nucl-th/0507033].

\bibitem{O'Connor:2007eb}
  E.~O'Connor, D.~Gazit, C.~J.~Horowitz, A.~Schwenk and N.~Barnea,
  Phys.\ Rev.\ C {\bf 75} (2007) 055803
  [nucl-th/0702044].

\bibitem{RoMuSchu1982}
      G.~R\"{o}pke, L.~M\"{u}nchow and H.~Schulz,
      Nucl.\ Phys.\ A {\bf 379} (1982) 536.

\bibitem{RoSchmMuSchu1983}
      G.~R\"{o}pke, M.~Schmidt, L.~M\"{u}nchow and H.~Schulz,
      Nucl.\ Phys.\ A {\bf 399} (1983) 587.

\bibitem{Hempel:2013tfa}
  M.~Hempel, V.~Dexheimer, S.~Schramm and I.~Iosilevskiy,
  Phys.\ Rev.\ C {\bf 88} (2013) 1,  014906
  [arXiv:1302.2835 [nucl-th]].

\bibitem{Bender:2003jk}
  M.~Bender, P.~H.~Heenen and P.~G.~Reinhard,
  Rev.\ Mod.\ Phys.\  {\bf 75} (2003) 121.

\bibitem{Typel:2009sy}
  S.~Typel, G.~R\"{o}pke, T.~Kl\"{a}hn, D.~Blaschke and H.~H.~Wolter,
  Phys.\ Rev.\ C {\bf 81} (2010) 015803
  [arXiv:0908.2344 [nucl-th]].

\bibitem{Voskresenskaya:2012np}
  M.~D.~Voskresenskaya and S.~Typel,
  Nucl.\ Phys.\ A {\bf 887} (2012) 42
  [arXiv:1201.1078 [nucl-th]].

\bibitem{Typel:2012esa}
  S.~Typel,
  AIP Conf.\ Proc.\  {\bf 1520} (2013) 68.

\bibitem{Typel:2013zna}
  S.~Typel, H.~H.~Wolter, G.~R\"{o}pke and D.~Blaschke,
  Eur.\ Phys.\ J.\ A {\bf 50} (2014) 17
  [arXiv:1309.6934 [nucl-th]].

\bibitem{Typel:1999yq}
  S.~Typel and H.~H.~Wolter,
  Nucl.\ Phys.\ A {\bf 656} (1999) 331.

\bibitem{Duflo:1994tr}
  J.~Duflo and A.~P.~Zuker,
  Phys.\ Rev.\ C {\bf 52} (1995) 23
  [nucl-th/9404019].

\bibitem{Tews:2012fj}
  I.~Tews, T.~Kr\"{u}ger, K.~Hebeler and A.~Schwenk,
  Phys.\ Rev.\ Lett.\  {\bf 110} (2013) 3,  032504
  [arXiv:1206.0025 [nucl-th]].

\bibitem{Kruger:2013kua}
  T.~Kr\"{u}ger, I.~Tews, K.~Hebeler and A.~Schwenk,
  Phys.\ Rev.\ C {\bf 88} (2013) 025802
  [arXiv:1304.2212 [nucl-th]].

\bibitem{Hempel:2011kh}
  M.~Hempel, J.~Schaffner-Bielich, S.~Typel and G.~R\"{o}pke,
  Phys.\ Rev.\ C {\bf 84} (2011) 055804
  [arXiv:1109.0252 [nucl-th]].

\bibitem{Li:2014oda}
  B.~A.~Li, A.~Ramos, G.~Verde and I.~Vida\~{n}a,
  Eur.\ Phys.\ J.\ A {\bf 50} (2014) 9.

\bibitem{Brown:2000pd}
  B.~A.~Brown,
  Phys.\ Rev.\ Lett.\  {\bf 85} (2000) 5296.

\bibitem{TyBr2001}
      S.~Typel and B.~Alex~Brown,
      Phys.\ Rev.\ C {\bf 64} (2001) 027302.

\bibitem{Vinas:2013hua}
  X.~Vi\~{n}as, M.~Centelles, X.~Roca-Maza and M.~Warda,
  Eur.\ Phys.\ J.\ A {\bf 50} (2014) 27
  [arXiv:1308.1008 [nucl-th]].

\bibitem{Typel:2014tqa}
  S.~Typel,
  Phys.\ Rev.\ C {\bf 89} (2014) 6,  064321
  [arXiv:1403.2851 [nucl-th]].

\bibitem{Aumann2014}
 T.\ Aumann and T.\ Uesaka, 
 private communication (2014).

\bibitem{Hofmann:2000vz}
  F.~Hofmann, C.~M.~Keil and H.~Lenske,
  Phys.\ Rev.\ C {\bf 64} (2001) 034314
  [nucl-th/0007050].

\bibitem{Antic:2015tga}
  S.~Anti\'{c} and S.~Typel,
  Nucl.\ Phys.\ A {\bf 938} (2015) 92
  [arXiv:1501.07393 [nucl-th]].

\end{thebibliography}
\end{document}